\documentclass[]{sig-alternate}
\pdfoutput=1
\usepackage{graphicx, xspace, subfigure,outlines,url,grffile,enumerate,listings,appendix,balance}
\usepackage[normalem]{ulem}
\usepackage{units}
\usepackage{paralist}
\usepackage{amsmath}
\usepackage{amsfonts}
\usepackage{xcolor}
\usepackage{changebar}
\usepackage[scriptsize]{caption}
\usepackage{microtype}
\usepackage{cite}
\usepackage{times}
\usepackage{mathptmx}
\makeatletter
\newcommand\subparagraph{%
  \@startsection{subparagraph}{5}
  {\parindent}
  {3.25ex \@plus 1ex \@minus .2ex}
  {-1em}
  {\normalfont\normalsize\bfseries}}
\makeatother
\usepackage[compact]{titlesec}
\let\subparagraph\relax

\usepackage{enumitem}
\setitemize{itemsep=1pt,topsep=1pt,parsep=1pt,partopsep=1pt}
\newcommand{\bi}{\begin{itemize}}
\newcommand{\ei}{\end{itemize}}

\newcommand{\eg}{{\it e.g.,}\xspace}
\newcommand{\ie}{{\it i.e.,}\xspace}
\newcommand\eat[1]{}
\newcommand\paragraphb[1]{\noindent{\bf #1}}
\newcommand\secref[1]{\S\ref{#1}}

\newcommand{\allnotes}[1]{}
\renewcommand{\allnotes}[1]{\textit{#1}}
\newcommand{\fixme}[1]{\allnotes{\bf\textcolor{red}{[#1]}}}

\newcommand{\noteori}[1]{\allnotes{\textcolor{green}{[Ori: #1]}}}
\newcommand{\notemooly}[1]{\allnotes{\textcolor{purple}{[Mooly: #1]}}}
\newcommand{\notekaterina}[1]{\allnotes{\textcolor{gray}{[Katerina: #1]}}}
\newcommand{\notepanda}[1]{\allnotes{\textcolor{cyan}{[Panda: #1]}}}
\newcommand{\katerinanote}[1]{\allnotes{\textcolor{gray}{[Katerina: #1]}}}
\newcommand{\name}{VMN\xspace}

\def\squarebox#1{\hbox to #1{\hfill\vbox to #1{\vfill}}}

\DeclareSymbolFont{extraup}{U}{zavm}{m}{n}
\DeclareMathSymbol{\vardiamond}{\mathalpha}{extraup}{87}
\newcommand{\link}[1]{\langle #1\rangle}

\usepackage{framed}

\usepackage{hyperref}

\colorlet{shadecolor}{gray!25}   
{\normalsize \endMakeFramed}
\title{Verifying Reachability in Networks with Mutable Datapaths}
 \author{\rm Aurojit Panda$^*$\hspace{-1em} \and \rm Ori Lahav$^\dagger$\hspace{-1em} \and \rm Katerina Argyraki$^\ddagger$\hspace{-1em} \and \rm Mooly Sagiv$^\diamondsuit$\hspace{-1em} \and \rm Scott Shenker$^{*\spadesuit}$\and
 {\rm $^*$UC Berkeley $^\dagger$MPI-SWS $^\ddagger$EPFL $^\diamondsuit$TAU $^\spadesuit$ICSI}}

\date{}
\begin{document}
\setlength{\abovedisplayskip}{0.2ex}
\setlength{\belowdisplayskip}{0.2ex}
\setlength{\abovedisplayshortskip}{0.2ex}
\setlength{\belowdisplayshortskip}{0.2ex}

\vspace{-1.5in}
\maketitle
\vspace{-1.8in}
\vskip -2.5em
\begin{abstract}
Recent work has made great progress in verifying the forwarding correctness of networks~\cite{mai2011debugging,khurshid2012veriflow,kazemian2012header,kazemian2013real}. However, these approaches cannot be used to verify networks containing middleboxes, such as caches and firewalls, whose forwarding behavior depends on previously observed traffic. We explore how to verify reachability properties for networks that include such ``mutable datapath'' elements. We want our verification results to hold not just for the given network, but also in the presence of failures. The main challenge lies in scaling the approach to handle large and complicated networks, We address by developing and leveraging the concept of slices, which allow network-wide verification to only require analyzing small portions of the network. We show that with slices the time required to verify an invariant on many production networks is independent of the size of the network itself.
\end{abstract}

\section{Introduction}
\label{sec:introduction}

Perhaps lulled into a sense of complacency because of the Internet's best-effort delivery model, which makes no explicit promises about network behavior, network operators have long relied on best-guess configurations
and a ``we'll fix it when it breaks'' operational attitude. However, as networking matures as a field, and institutions increasingly rely on networks to provide reachability, isolation, and other behavioral constraints, there is growing interest in developing rigorous verification tools that can ensure that these constraints will be enforced by the network configuration. The first generation of such tools 
-- Anteater \cite{mai2011debugging}, Veriflow \cite{khurshid2012veriflow}, and HSA \cite{kazemian2012header,kazemian2013real} -- provide highly efficient (in fact, near real-time) checking of reachability (and, conversely isolation) properties and detect anomalies such as loops and black holes. This technical advance represents an invaluable step forward for networking.

These verification tools assume that the forwarding behavior is set by the control plane, and not altered by the traffic, so verification needs to be invoked only when the control plane alters routing entries. This approach is entirely sufficient for networks of routers, which is obviously an important use case. However, modern networks contain more than routers. 

Most networks contain switches whose learning behavior renders their forwarding behavior dependent on the traffic they have seen. More generally, most networks 
also contain middleboxes, and middleboxes often have forwarding behavior that depends on the observed traffic. For instance, firewalls often rely on outbound 
``hole-punching'' to allow hosts to establish flows to the outside world, and content caches forward differently based on whether they have previously cached the
desired content. We refer to network elements whose forwarding behavior can be altered by datapath activity as having a ``mutable datapath'' (in contrast with
static datapaths whose behavior is fixed until the control plane intervenes), and additional examples of such elements include WAN optimizers,
deep-packet-inspection boxes, and load balancers. In short, the behavior of a static datapath is only a function of its configuration, while the behavior of a
mutable datapath also depends on the entire packet history that it has seen. 

\eat{\notepanda{Furthermore, for many middleboxes, in contrast to switches, losing state due
to failures can result in incorrect or unsafe behavior, further complicating network configuration.}

\notepanda{We might want to add something about the implications of adding middleboxes to network: (a) Need to make sure that combining middleboxes in a pipeline (composing) requires that one also account for how the behavior will change due to state; (b) Planning for failures is hard since failures lead to state loss (not as simple as just routing around the failure).}}

While classical networking often treats middleboxes as an unfortunate and rare occurrence in networks, in reality middleboxes are the most viable way to incrementally deploy new network functionality. Operators have turned to middleboxes to such a great extent that a recent study \cite{sherry2012making} of fifty-seven enterprise networks revealed that these networks are roughly equally divided between routers, switches and middleboxes. Thus, roughly two-thirds of the forwarding boxes in enterprise networks can have mutable datapaths that would not conform to the models used in the recently developed network verification tools. In addition, the rise of Network Function Virtualization (NFV)~\cite{nfv}, in which physical middleboxes are replaced by their virtual counterparts, makes it easier to deploy additional middleboxes without changes in the physical infrastructure. Thus, we must reconcile ourselves to the fact that many networks will have substantial numbers of elements with mutable datapaths (and hereafter, when referring to such elements we will call them middleboxes).
\eat{\notepanda{Somehow indicate that we cannot solve the Google outage, and are merely offering it as proof that middleboxes are an issue.} Moreover, not only are middleboxes prevalent, but they are often responsible for network problems. In December 2012 a misconfiguration in Google's load balancers resulted in a several minute outage for GMail and other Google Services~\cite{googleoutage}. \fixme{While Google's service outage was caused by load balancer misconfiguration resulting in increased load on servers, and was not an isolation invariant (which we focus on in this paper), \ldots.}}  Moreover, not only are middleboxes prevalent, but they are often responsible for network problems. A recent two year study~\cite{potharaju2013demystifying} of a provider found that middleboxes played a role in 43\% of their failure incidents, and between 4\% and 15\% of these failures were the result of middlebox misconfiguration. \eat{Thus, middleboxes are a significant cause of network problems, and are not supported by existing verification tools.}

The goal of this paper is to extend the notion of verification to networks containing mutable datapaths, so that such middlebox misconfiguration problems can be prevented.\footnote{While we are extending the class of networks -- to those including mutable datapaths -- we are {\em not} significantly expanding the class of invariants to be checked; just as in the earlier works, we are focusing on reachability and isolation.} Further, these techniques should ensure correctness even in the presence of failures, a requirement not addressed by any of the existing network verification tools. Our basic approach, which we call Verification for Middlebox Networks (\name) is simple: given a topology containing both mutable datapaths (\eg middleboxes) and static ones (\eg routers), we derive a logical formula that model the network as a whole. We then add logical formulas derived from the specified invariants so that an invariant holds if and only if there is no satisfying assignment for this set of logical formulas as a whole.  As described so far, this is a straightforward application of standard program verification techniques to networks.

However, na\"ively applied, this approach would not scale: middlebox code is complex,
and checking even simple invariants in modest-sized networks would be intractable. {\em Thus, our focus is on how to scale this approach to large networks.}
\name uses four techniques to scale verification: 

{\bf 1. Limited invariants}: Rather than deal with an arbitrary set of invariants, we focus on two specific categories that are the dominant concerns for network operators.  First, we look at invariants describing the set of middleboxes (more generally as a DAG of middleboxes) packets should flow through (\eg all http traffic should pass through a firewall then a cache); we call these {\it pipeline} invariants. Second, we also consider invariants that address reachability/isolation between hosts (at the packet or content level), such as packets from host A should not reach host B (and hereafter we will call these reachability invariants). Our contributions relate to verifying reachability invariants, and we rely on established techniques to verify pipeline invariants. 

{\bf 2. Simple high-level middlebox models}: One approach to verifying networks with middleboxes would be to use their full implementation to determine their behavior. This is infeasible for two reasons: (i) most commonly deployed middleboxes are proprietary, and we do not have access to their code, and (ii) model checking even one such box for even the simplest invariants would not scale.\footnote{This assertion does not contradict the results in \cite{dobrescu2014software}, which we discuss later in the paper.} Therefore, we model middleboxes using a simple abstract forwarding model and a set of abstract packet classes used by this model. We do not model the packet classification algorithm in middleboxes, and instead rely on an oracle to classify packets.\footnote{A third reason we do not apply model checking to the full implementation, which we discuss in the next section, is that there is a semantic mismatch between raw middlebox code and operator-specified invariants, which are described in terms of basic abstractions. In fact, this mismatch is what led us to the combination of an Oracle and an abstract model.} The forwarding models can typically be derived from a general description of the middlebox's behavior and can be easily analyzed using standard techniques.

{\bf 3. Modularized network models}: Networks contain elements with static datapaths and elements with mutable datapaths. Rather than consider them all within the one verification framework, which would overburden a system already having trouble scaling, we treat the two separately: it is the job of the static datapath elements to satisfy the pipeline invariants (that is, to carry packets through the appropriate set of middleboxes), which we can analyze using existing verification tools; and it is the job of the processing pipeline to enforce reachability invariants, and that is where we focus our attention.  Thus, our resulting system is a hybrid of current static-datapath verification tools and our newly-proposed tool for mutable datapaths. 

{\bf 4. Symmetries and Common Cases} If middleboxes are \emph{flow-parallel} or \emph{origin-agnostic} (both to be defined later, and most middleboxes fall into one or both categories), we can perform network-wide verification by examining only a small portion of the network. This allows us to scale verification of a single invariant to large networks. Furthermore, operational networks exhibit a great deal of symmetry in how they are structured and in the policies they enforce. We exploit this symmetry to reduce the total time taken for verifying all invariants in the network. The combination of these two observations allows us to verify the correctness of large networks in a few seconds.
\smallskip

Prior work, particularly Buzz~\cite{fayaz2014buzz}, have described network tools (akin to ATPG~\cite{zeng2012automatic}) that can be applied to networks with middleboxes; they generate test packets that (when sent) efficiently explore whether or not invariants are violated. In contrast, \name provides mechanisms for verifying (akin to HSA~\cite{kazemian2013real}) reachability invariants in networks with middleboxes. Our contributions also include slicing, which allows verification to scale to arbitrarily large networks (through the use of \emph{slices}), and checking whether invariant hold during failures.
We provide a more detailed comparison between \name and other systems in \S\ref{sec:related}.

In the next section, we discuss all four of these steps more formally, and then in \secref{sec:system} we provide an overview of \name. We discuss our strategy for performing
verification on smaller, fixed size subnetworks for scalability in \secref{sec:scalability}. In \secref{sec:eval} we evaluate \name by verifying invariants for a variety of real-world scenarios.
Finally we conclude in \secref{sec:related} and \secref{sec:conclusion} with a discussion of related work and a brief summary.

\eat{Furthermore, SymNet \cite{stoenescu2013symnet} also observed that dealing with middleboxes is essential when analyzing networks and uses symbolic execution to extend HSA to deal with hole-punching firewalls. \name relies on different techniques and our treatment of state is far more general, extending to content caches and other more complex stateful middleboxes.} 

\section{Our Approach}
\label{sec:modeling}

In this section we provide an overview of our approach.

\subsection{Invariants}
\label{sec:modeling:invariants}
The purpose of verification is to test whether some properties ({\em invariants}) hold for a given network, where a network is defined by both its {\em topology} (location of routers/switches and middleboxes) and its {\em configuration} (routing tables and middlebox settings, including how both routers and middleboxes respond to failures). The verification techniques employed necessarily depend on the nature of the invariants to be checked. Thus, to understand the verification problem we are tackling, we must first specify the class of invariants we consider.

Verifying arbitrary invariants for networks with mutable datapaths is undecidable\footnote{Middleboxes render the network Turing complete, so verifying certain invariants is equivalent to solving the halting problem.}~\cite{TACAS}, which is why we focus more narrowly on pipeline and reachability invariants for various classes of packets that are defined in terms of hosts, users, source/destination addresses, 
ingress/egress ports, flows, content, application, whether or not a packet is ``malicious'' (as decided, for example, by a DPI box), whether or not a packet belonging to this flow has been seen before, and other concepts. These invariants can refer to the current network configuration, or be predicated on one or more failures in the network (\ie one can insist that reachability or pipeline invariant not only hold for the current network, but for all single failures in that network).

\subsection{High-Level Middlebox Models} 
Some invariants are defined in terms of packet classes based only on {\em intrinsic} information --- such as physical ports and header fields --- that can
be precisely defined by network operators. However, operators frequently rely on invariants defined in terms of higher-level abstractions --- such as what
user a flow belongs to, what application sent a flow, whether the packet or flow is malicious, etc. --- that depend on other context (for instance authentication services
implemented outside of a particular middlebox), and can often not be precisely defined (whether a packet is malicious depends on the current set of CVEs~\cite{cve} and various heuristics). Using  high-level abstractions enables operators to express their intent (such as to drop all malicious traffic, or drop all traffic from a given user) without having to specify, within the invariant itself, the precise mechanisms used to define those abstractions. For example, an operator may wish to drop all Skype traffic, but does not know (or care) about the precise mechanisms an application-level firewall uses to identify such traffic. Therefore, to reflect how things are currently done when configuring middleboxes or expressing policies (\eg Congress \cite{congress}), we allow invariants to be defined in terms of these higher-level abstractions.  

{\em How do we model middleboxes when their code is necessarily in terms of intrinsic information but the invariants can be
in terms of higher-level abstractions?}  We choose to describe middleboxes in two stages: we start with a simple abstract
forwarding model coupled with a set of abstract packet classes (is a packet malicious, is it a part of a Skype flow, etc.) and any requirements
for classifying packets into this type (\eg, to determine whether a packet belong to a Skype connection the middlebox implementation needs to see all packets in a flow). \name then augments this model by adding a \emph{classification oracle} which is responsible for classifying packets into these packet classes. The abstract forwarding model describes how the middlebox operates on a packet (\ie whether and where it is forwarded, and whether the header is rewritten) given the
intrinsic information (\ie packet header and port information), the entire packet history (since middlebox datapaths can be mutable, their behavior can depend on packet history), and the abstractions it is assigned by the oracle (is the packet part of a Skype flow?).  The \emph{type} of middlebox (\eg firewall or load balancer) describes the basic behaviors supported by the middlebox (for
instance, which abstractions it supports, whether it rewrites packet headers, etc.), but the \emph{configuration} of the
middlebox dictates to which class of packets these behaviors are applied. 

The task of verification that we address in this paper is whether the invariants are obeyed {\em assuming that middlebox implementation can correctly classify packets into abstract types.} That is, we verify network and middlebox configuration, not the implementation of the abstractions. Clearly, if a middlebox cannot correctly classify traffic (\eg Skype traffic) an invariant might not hold. However, this error is not caused by network configuration but rather by bugs in the middlebox implementation. While verifying middlebox implementation is an important task, it is beyond the scope of this paper. Here we focus solely on whether the overall network configuration upholds invariants assuming middleboxes are correctly implemented.

\subsection{Modularized Network Models}
The tools developed for static datapaths~\cite{kazemian2012header,kazemian2013real} not only verify invariants, but also summarize the behavior of such networks as transfer functions. 
A transfer function for a network of static datapaths maps an incoming packet on a physical port at the edge of the network to one or more outgoing packets (perhaps with rewritten header fields) departing out one or more physical ports at the edge of the network. Thus, we can consider our network as a set of elements with mutable datapaths tied together by transfer functions which represent the behavior of the static datapath portion of the network.\footnote{If the static datapaths lead to a loop for a particular packet, we raise an exception (so the network operator is aware of it) and treat the packet as dropped.} When a failure occurs, routing in the static datapaths will change, producing a new transfer function. Rather than model the details of the routing algorithm, we assume we are given a function mapping failure conditions to these new transfer functions (\eg a list of backup paths taken in response to failures). \eat{Therefore, for a given network we consider a set of transfer functions and a function mapping each failure condition (\ie the set of failed middleboxes) with the transfer function to be used.}

The glue provided by these transfer functions is precisely what is responsible for enforcing pipeline invariants. In its simplest incarnation, a pipeline invariant takes the form: {\it all incoming packets with a certain class of headers must pass through the sequence of middleboxes $mb_1,\ mb_2,\ mb_3,\ ...$ before being delivered to the intended destination}. More complicated pipeline invariants involve a DAG of middleboxes and specify the appropriate branching at each step (\eg all \texttt{http} packets leaving the firewall go to the load balancer, while all other traffic goes directly to the destination). Note that these invariants could refer to physical instances of middleboxes (\eg packets must traverse this particular middlebox) or
a class of middleboxes (\eg packets must traverse a firewall).  In what follows, we will assume that all packets belonging to the same flow are processed by the same physical pipeline (this can be easily enforced in existing networks).


Once we have decomposed the network into a set of middleboxes connected by transfer functions, checking these pipeline invariants is straightforward using existing network verification tools.  The reachability invariants are more difficult, as we discuss next.

\subsection{Scaling Verification}
For moderately sized networks, we can use the techniques discussed above to generate logical formulas modeling network behavior, and other formulas corresponding to the invariants that need to be verified. We can then use an SMT solver (\eg Z3~\cite{de2008z3}) to check if the invariants hold for the provided network or not. However as the scale of the network increases, the SMT solver has to account for an exponentially larger state space, slowing down or preventing verification.

Inspired by compositional verification~\cite{tse:MisraC81, ifip:Jones83} (which allows the results from verifying components of a program to be combined to reason about the program as a whole) we have identified a general class of subnetworks, which we refer to as {\em slices}, where any invariant that only references middleboxes or endhosts contained in a slice holds for the entire network if and only if it holds for the slice. Therefore, we can scale network verification if given a network and an invariant, we can find a slice whose size is independent of the size of the network and which contains all middleboxes or endhosts referenced by the invariant. 

While slices help us rapidly verify that an individual invariant holds in a network, a network might implement several invariants. Prior work~\cite{symmetryandsurgery} has observed that many operational networks have symmetric topologies and policies. Since the proof generated while verifying that a particular invariant holds also applies to all other symmetric invariants, this greatly reduces the time taken to verify the behavior of a network allowing us to scale verification to extremely large operational networks.

\newcommand{\globally}[1]{\Box{#1}}
\newcommand{\past}[1]{\vardiamond{#1}}
\newcommand{\rcv}{rcv}
\newcommand{\snd}{snd}
\newcommand{\src}{src}
\newcommand{\bdy}{body}
\newcommand{\dst}{dst}
\newcommand{\fail}{fail}

\section{System Design}
\label{sec:system}
We model network behavior in discrete timesteps. During each timestep a previously sent packet can be delivered to a node (middlebox or host), a host can generate a new packet that enters the network, a middlebox can process a previously received packet, a failure can occur, or a previously failed node can recover. We do not attempt to model the \emph{likely} order of these various events, but instead consider all such orders in search of invariant violations. To do so, we invoke a \emph{scheduling oracle} that assigns a single event to each timestep, subject to the constraint that ordering (both for receiving and processing) is maintained for packets sent on the same link. \eat{Give this model, \name verifies the invariant by searching for a schedule, representing a sequence of packets and failure events that result in the provided invariant being violated. The invariant holds if and only if no such sequence can be found.}

\subsection{Overview}
\name accepts as input a set of middlebox models (\S\ref{sec:system:modeling}), and a set of transfer functions (\S\ref{sec:system:transfer}) with a mapping between failure conditions and transfer functions. \eat{We perform this search by adding a \emph{scheduling} oracle, a \emph{classification} oracle and a \emph{failure} oracle to the network and finding cases where an oracle's choices can result in an invariant violation. 

The  The \emph{classification oracle} is responsible for deciding if a packet belongs to a particular abstract packet class (\eg Skype from the previous section). The classification oracle respects any user specified constraints on classification. Finally, the \emph{failure oracle} is responsible for scheduling middlebox failures, and decides the set (possibly empty) of middleboxes that are reported to have failed in this timestep. The failure oracle influences both middlebox behavior and the transfer function used at each timestep.}
\name builds on Z3~\cite{z3} a state of the art SMT solver. SMT solvers accept as input a set of boolean formulae expressed in terms of a set of variables, and then either find an assignment for the variables such that the conjunction of these formulae is true (a satisfying assignment) or a proof that no such assignment exists. In this section we present our technique for converting a network and set of invariants to a set of axioms, and producing a logical formulae including these axioms. We can then use Z3 to check satisfiability for these formulae, where a satisfying assignment indicates that an invariant is violated by the network.

In \name middleboxes and networks are modeled using quantified formula, which are axioms describing how received packets are treated, while the classification and scheduling oracles are modeled using variables. In addition to the network model and oracles, we also provide Z3 with the negation of the invariant, which we specify in terms of a set of packets. Finding a satisfiable assignment to these formulae is equivalent to Z3 finding a set of oracle behaviors that result in the invariant being violated, and proving the formulae unsatisfiable is equivalent to showing that no oracular behavior can result in the invariants being violated.

The search problem solved by SMT solvers is undecidable in general, and they rely on heuristics and timeouts to ensure their search procedure terminates (in which case they are unable to determine if a problem is satisfiable or not). Na\"ively generating middlebox and network formulae might yield formulae in an undecidable logic, preventing successful verification. Therefore, one of the core contribution of \name lies in producing logical formulae for a wide range of networks (and a variety of middleboxes) and invariants in a weak
logic that is expressible in Z3 and for which verification succeeds in practice.\eat{\footnote{We provide a detailed discussion of decidability in Appendix~\ref{sec:tractability}.}}  In the rest of this section we explore these models and formula in greater depth.

\eat{
\begin{table}[tbp]
\centering
\begin{tabular}{|l|p{0.67\columnwidth}|}
\hline
Symbol & Meaning\\
\hline
\hline
\multicolumn{2}{|c|}{Events}\\
\hline
$\rcv(d, s, p)$ & Destination $d$ receives packet $p$ from source $s$.\\
\hline
$\snd(s, d, p)$ & Source $s$ sends packet $p$ to destination $d$.\\
\hline
\multicolumn{2}{|c|}{Logical Operators}\\
\hline
$\globally{P}$ & Condition $P$ holds at all times.\\
\hline
$\past{P}$ & Event $P$ occurred in the past.\\
\hline
$\neg P$ & Condition $P$ does not hold (or event $P$ does not occur).\\
\hline
$P_1 \land P_2$ & Both conditions $P_1$ and $P_2$ hold.\\
\hline
$P_1 \lor P_2$ & Either condition $P_1$ or $P_2$ holds.\\
\hline
\end{tabular}
\caption{Logical symbols and their interpretation.}
\label{tab:ltl}
\end{table}
}
\vspace{-0.03in}
\subsection{Notation}
\label{sec:system:notation}
We begin by presenting the notation used in this section. We express our models and invariants using a simplified form of linear temporal logic (LTL)~\cite{LichtensteinPZ85} of events, with past operators. We restrict ourselves to safety properties, and hence only need to model events occurring in the past or events that hold globally for all of time. We use LTL for ease of presentation; \name automatically converts LTL formulas into first-order logic (as required by Z3) by explicitly quantifying over time.

\sloppy
Our formulas are expressed in terms of three events: $\snd(s, d, p)$, the event where a \emph{node} (end host, switch or middlebox) $s$ sends \emph{packet} $p$ to
\emph{node} $d$; and $\rcv(d, s, p)$, the event where a node $d$ receives a packet $p$ from node $s$, and $\fail(n)$, the event where a node $n$ has failed. Each event happens at a timestep and logical formulas can refer either to events that occurred in the past (represented using $\past{}$) or properties that hold at all times (represented using $\globally{}$). For example, 
\begin{align*}
\forall d,s,p:\ \globally (\rcv(d, s, p) \implies \past{\snd(s, d, p)})
\end{align*}
says that at all times, any packet $p$ received by node $d$ from node $s$ must have been sent by $s$ in the past.

Similarly, \begin{align*}
\forall p:\ \globally{\neg \rcv(d, s, p)}
\end{align*}
indicates that $d$ will never receive any packet from $s$. 

Header fields and abstract packet classes are represented using functions, \eg $\src(p)$ and $\dst(p)$ represent the source and destination address for packet $p$, and $skype?(p)$ returns true if and only if $p$ belongs to a Skype session. 

\vspace{-0.03in}
\subsection{Reachability Invariants}
\label{sec:system:invariants}
Reachability invariants can be be generally specifies as:
\begin{align*}
\forall n, p:\ \globally{\neg (\rcv(d, n, p) \land predicate(p))},
\end{align*}
which says that node $d$ should never receive a packet $p$ that matches $predicate(p)$. The $predicate$ can be expressed
in terms of packet-header fields, abstract packet classes and past events, this allows us to express a wide variety of network properties as reachability invariants, \eg:
\begin{asparaitem}
\item Simple isolation: node $d$ should never receive a packet with source address $s$. We express this invariant using the $\src$ function, which extracts the source IP address from the packet header:
\[
\forall n, p: \globally{ \neg (\rcv(d, n, p) \land \src(p) = s)}.
\]
\item Flow isolation: node $d$ can only receive packets from $s$ if they belong to a previously established flow. We express this invariant using the $flow$ function, which computes a flow identifier based on the packet header:
 \begin{align*}
\forall n_0, p_0,& n_1, p_1: \globally{} \neg (\rcv(d, n_0, p_0)\land \src(p_0) = s \land\\
&\neg(\past{\snd(d, n_1, p_1) \land flow(p_1) = flow(p_0)})).
\end{align*}
\item Data isolation: node $d$ cannot access any data originating at server $s$, this requires that $d$ should not access data either by directly contacting $s$ or indirectly through network elements such as content cache. We express this invariant using an $origin$ function, that computes the origin of a packet's data based on the packet header (\eg using the \texttt{x-http-forwarded-for} field in HTTP):
\[
\forall n, p: \globally{ \neg (\rcv(d, n, p) \land origin(p) = s)}.
\]
\eat{Note, that in general during verification we assume end hosts are well-behaved, \ie we do not verify cases where an endhost proxies content to another endhost. Rather, the data isolation invariant here focuses on cases where caches or other network elements cache data.}
\end{asparaitem}
In addition, \name can verify several other invariants, including whether packets traverse a certain link or middlebox.

\begin{lstlisting}[caption=Model for a learning firewall,
                   label=code:lfw,float
                   ]
@FailClosed
class LearningFirewall (acl: Set[(Address, Address)]) {
  val established : Set[Flow]
  def model (p: Packet) = {
    when established.contains(flow(p)) =>
      forward (Seq(p))
    when acl.contains((p.src, p.dest)) =>
      established += flow(p)
      forward(Seq(p))
    _ =>
      forward(Seq.empty)
  }
}
\end{lstlisting}

\vspace{-0.03in}
\subsection{Modeling Middleboxes}
\label{sec:system:modeling}
\eat{Middlebox models in \name are specified by middlebox providers or network operators using a high-level language. The specification include both a set of \emph{high-level abstract classifiers} and an abstract forwarding model. High-level classifiers are specified by 
providing the type of input(s) required by the classifier and the type of output provided; and the forwarding model is specified using a loop-free, event-driven language based on Scala syntax.
Our middlebox models depend only on the type of the middlebox (where the type specifies both the middlebox behavior and the high-level abstractions implemented by the middlebox), and not on its implementation or placement.
Since there are only a limited number of middlebox types in wide deployment~\cite{rfc3234}, we believe that in a majority of cases users of our tool can reuse existing models.}

Middleboxes in \name are modeled using a high-level loop-free, event driven language. Restricting the language so it is loop free allows us to ensure that middlebox models are expressible in first-order logic (and can serve as input to Z3). We use the event-driven structure to translate this code to logical formulae (axioms) encoding middlebox behavior.

Middlebox models are specified as a class containing the \emph{abstract packet classes} the middlebox depends on, its \emph{forwarding model}, and its \emph{failure behavior}. \emph{Abstract packet classes} are specified as a set of function prototypes. Optionally, models can also specify input constraint that must be met for the implementation to correctly identify that a packet belongs to a particular class, and output constraints restricting the set of classes a packet can belong to. For example, an application firewall might specify an abstract packet class for each application (\eg \texttt{skype?},~\texttt{jabber?}), specify that correct identification requires all packets in a flow to go through the same middlebox instance, and specify that a packet can belong to at most one application class (a packet cannot be both a Skype packet and a Jabber packet). Middlebox \emph{forwarding models} are specified as functions which dictate how packets are modified and whether they are forwarded or dropped. Complex packet modification, \eg encryption or compression, are modeled as replacing the appropriate packet header field (or payload) with a random value, this provides sufficient fidelity for checking reachability invariants. Finally, middlebox failure behavior is specified either explicitly in the forwarding model, using the \texttt{fail} predicate provided by \name, or implicitly by specifying whether a middlebox fails-closed (\ie packets are dropped during middlebox failures) or fails-open (\ie all received packets are forwarded unmodified during failure). We provide examples of such specification below.

Listing~\ref{code:lfw} shows the specification for a stateful firewall. The model accepts a set of ACLs (\texttt{acl} on Line 2) as configuration, and maintains flow state (in the \texttt{established} 
variable defined on Line 3). On receiving a packet from an established flow, the firewall forwards the packet (Line 5 and 6), otherwise it checks to see if the packet is permitted by the 
firewall configuration (Line 6--10). The firewall forwards the packet if permitted and drops it otherwise. The \texttt{@FailClosed} annotation on line 1 indicates that the firewall fails closed, and packets are dropped during failure.

Similarly, Listing~\ref{code:nat} shows the model for a NAT. In this example we explicitly model failure behavior (Line 6), and the NAT drops packets when failed. We also modify the packet's source and destination port (Line 10--11, 14--15, and 20--21) as a part of the forwarding behavior. We assign ports to new flows at random by calling the \texttt{remapped\_port} (line 2) method.

\begin{lstlisting}[caption=Model for a NAT,
                   label=code:nat,float]
class NAT (nat_address: Address){
 abstract remapped_port (p: Packet): int
 val active : Map[Flow, int]
 val reverse : Map[port, (Address, int)]
 def model (p: Packet) = {
   when  fail(this) =>
     forward(Seq.empty)
   dst(p) == nat_address =>
     (dst, port) = reverse[dst_port(p)];
     dst(p) = dst;
     dst_port(p) = port;
     forward(Seq(p))
   active.contains(flow(p)) =>
     src(p) = nat_address;
     src_port(p) = active(flow(p));
     forward(Seq(p))
   _ =>
     address = src(p);
     port = src_port(p)
     src(p) = nat_address;
     src_port(p) = remapped_port(p);
     active(flow(p)) = src_port(p);
     reverse(src_port(p)) = (address, port);
     forward(Seq(p))
 }
}
\end{lstlisting}

\name translates these high-level specifications into a set of parametrized axioms (the parameters allow more than one instance of
the same middlebox to be used in a network). For instance, Listing~\ref{code:lfw} results in the following axioms:

\begin{flalign*} 
    \textbf{established}(flow(p)) &\implies (\vardiamond ((\neg fail(\textbf{f})) \land (\vardiamond \rcv(\textbf{f}, p))))&\\
    &\land \textbf{acl}(\src(p), \dst(p))&
\end{flalign*}
\begin{flalign*}
    send(\textbf{f}, p) &\implies (\vardiamond \rcv(\textbf{f}, p))&\\
    &\land (\textbf{acl}(\src(p), \dst(p))& \\
    &\lor \textbf{established}(flow(p)))&
\end{flalign*}
The bold-faced terms in this axiom are parameters: for each stateful firewall that appears in 
a network, \name adds a new axiom by replacing the terms $\textbf{f}$, $\textbf{acl}$ and $\textbf{established}$ with 
a new instance specific term. The first axiom says that the $\textbf{established}$ set contains a flow if a packet permitted by the firewall policy ($\textbf{acl}$) has been received by $\textbf{f}$ since it last failed. The second one states that 
packets sent by $\textbf{f}$ must have been previously received by it, and are either pr emitted by the $\textbf{acl}$'s for that firewall, or belong to a previously established connection. The axioms generated for the NAT are similar, and are elided due to space
constraints.

We require users of \name to provide middlebox models, however our models are at a high level and depend only on the type of the middlebox, not its placement or implementation details. Previous studies have found that only a limited number of middlebox types are widely deployed~\cite{rfc3234} in production networks\footnote{The existence of a limited number of middlebox types does not limit the number of deployed middleboxes. Networks commonly include several middleboxes belonging to the same type, this might be for resilience, improving network performance and to reduce the load on each middlebox.}, and we believe that in a majority of cases users of our tool can reuse existing models.

\eat{
The axiom in this case says that the $\textbf{established}$ set contains a flow if a packet permitted by the firewall policy ($\textbf{acl}$) was previously received by $\textbf{f}$. Any packets sent by $\textbf{f}$ must have previously been received, and is either permitted by $\textbf{acl}$ or 
belongs to a previously established connection.

To ensure that middlebox logic can be successfully converted to first-order logic we require that forwarding models meet a few requirements:
\begin{asparaenum}
\item They should be loop-free and process in a bounded number of steps.
\item They should only access local state. This is naturally true for existing middleboxes, which are physically distinct and must use the network to share
state.
\item They should be deterministic for a given packet and history. While this might not hold for some existing middleboxes (\eg load balancers which distribute
traffic according to a random distribution), we can find semantically equivalent deterministic versions for such middleboxes.
\end{asparaenum}

In practice we have found that these limits are met by most existing middleboxes (and it would be hard to build middleboxes with sufficient
performance whose forwarding behavior does not fit in these constraints). Further, these
restrictions allow us to express several existing middleboxes in a decidable logic.}

\subsection{Modeling Networks}
\label{sec:system:transfer}
\name uses \emph{transfer functions} which were previously developed by HSA~\cite{kazemian2012header} and VeriFlow~\cite{khurshid2012veriflow} to specify
a network's forwarding behavior during a particular failure scenario. The transfer function for a network is a function from a located packet (a packet augmented with the network port where it is located) to a set of located packets indicating where the packets are next sent. For example, the transfer function for a network with $3$ hosts $A$ (with IP address $a$), $B$ (with IP address $b$) and $C$ (with IP address $c$) is given by:
\[
f(p, port) \equiv \begin{cases}
(p, A) & \mbox{if } \dst(p) = a\\
(p, B) & \mbox{if } \dst(p) = b\\
(p, C) & \mbox{if } \dst(p) = c
\end{cases}
\]

\name assumes switch forwarding tables are static, however they might change depending on the failure scenario. Therefore, rather than accepting a single static network topology and configuration as input, \name accepts a topology and forwarding table corresponding
to each failure scenario. Given the topology and switch forwarding tables used by the network in a particular failure scenario, \name uses VeriFlow to compute a transfer function. In this computed transfer function, all ports correspond to either middleboxes or end-hosts, \ie the transfer function models the network as a set of end-hosts and middleboxes connected to a single switch. \name translates this transfer function to axioms by introducing a single pseudo-node ($\Omega$) representing the network, and deriving a set of axioms for this pseudo-node from the transfer function and failure scenario. For example, the previous transfer function is translated to the following axioms ($\fail(X)$ here represents the specified failure model).
\begin{align*}
\forall n, p: \globally{} \fail(X) &\land \ldots \snd(A, n, p) \implies n = \Omega\\
\forall n, p: \globally{} \fail(X) &\land \ldots \snd(\Omega, n, p) \land \dst(p) = a\\
&\implies n = A \land \past{\exists n': \rcv(n', \Omega, p)}
\end{align*}

VeriFlow (and HSA) can only produce transfer functions when the topology and forwarding table for a network are loop-free. \name therefore throws an exception
when a static forwarding loop is encountered. Not allowing loops in the forwarding logic is also important for allowing us to express network axioms in first-order logic
and helps ensure \name's verification process is decidable.

In addition to the axioms for middlebox behavior and network behavior, \name also adds axioms restricting the oracles' behavior, \eg we add axioms to ensure that any
packet delivery event scheduled by the scheduling oracle has a corresponding packet send event, and we ensure that new packets generated by hosts are well formed.

\eat{
VeriFlow and HSA can only produce transfer functions when the forwarding graph (which is the graph along which packets are forwarded) for a network
has no loops. We therefore throw an exception when this condition does not hold. More importantly, this means that the forwarding axioms used by
\name always represent acyclic paths along which packets are forwarded; this allows us to express network axioms in first-order logic and helps
with decidability.
In addition to these forwarding axioms, we also add a set of basic network axioms describing assumptions common to all networks. These
axioms include:
\begin{asparaitem}
\item Packet that are received must have been sent previously: $$\forall s, d, p:\ \globally{\rcv(d, s, p) \implies \past{\snd(s, d, p)}}$$
\item Packets cannot loop back to the sender: $$\forall s, p:\ \globally{\neg \snd(s, s, p)}$$
\item Packets are well formed, and the source address is distinct from the destination address:
$$\forall s, d, p:\ \globally{\snd(s, d, p) \implies \dst(p) \neq \src(p)}$$
\end{asparaitem}

Given a network (\ie topology, routing configuration, set of middlebox models and middlebox configurations), \name produces an
equivalent network axiom by taking the union of axioms for each middlebox, the network transfer function and basic network axioms.
}

\subsection{Limitations}
\label{sec:system:limitations}
To ensure verification is practical and tractable, our models of networks and middleboxes are necessarily abstract. This imposes some limitations for the results returned by \name. Firstly, we do not verify the classification logic in a middlebox implementation, our verification results are conditioned on packets being correctly classified by the middlebox. Therefore, our results might be wrong when classification logic is incorrectly implemented. This is a separable problem: \name does not obviate the need to verify and test individual middleboxes, it just provides a mechanism to verify the behavior of combined middleboxes. Providing tools to test or verify individual middleboxes is outside scope of our work.

Secondly, we do not have complete semantic information about abstract packet classes, and this can result in \name reporting false positives (\ie invariant violations) where none exist. For example, consider a network with two application specific firewalls, one that can identify Skype traffic, and another that can identify streaming audio services. A priori, \name has no information indicating that these packet classes are mutually exclusive, and will consider packets which meet both criterion when looking for invariant violations. This can be solved by augmenting \name's models with logical constraints encoding these assumptions, however we do not currently include such constraints.

Finally, our models do not contain semantics for complex packet modifications (\eg encryption, compression, etc.), and instead just change the affected packet to a random value. Similar to the previous cases this can also result in false positives in the same way as above.

Most of these limitations are fundamental, network verification without the use of abstractions is intractable, and is impractical with large models. Our choices allow us to provide useful verification, as shown \S\ref{sec:eval}, and in most practical cases we observed no false positives.
\section{Scaling Verification}
\label{sec:scalability}
Z3 (and other SMT solvers) rely on heuristics and timers to ensure satisfiability checking terminates, and cannot always prove (or disprove) satisfiability of large sets of formulae. The size of formulae produced by the techniques in \S\ref{sec:system} are proportional to the size of the network being verified, and therefore cannot be applied to large networks. Scaling therefore requires that the size of the formulae generated be independent of network size. We rely on \emph{network slices} as described here. \eat{for scaling; a network slice is a subnetwork whose behavior with respect to a set of invariants is the same as the entire network's behavior. Therefore the result of verifying a set of invariants on an appropriate slice naturally extend to the entire network. The size of a slice depends on both the invariants and network being verified, and slices smaller than the entire network cannot always be found. However, in practice, small (constant sized) slices can be used to verify reachability invariants in many real networks.}

We begin by providing an informal overview of network slices, a more formal description is available in our technical report.\footnote{Anonymized for double blind submission, we can provide proofs and other details on request.} Given a network $N$, a subnetwork $\Omega$ is a network formed from a subset of $N$'s nodes (middleboxes, hosts and switches) and links. All packets sent by hosts in subnetwork $\Omega$ and received by hosts in $\Omega$ are said to belong to $\Omega$. We say $\Omega$ is closed under forwarding if and only if all packets belonging to $\Omega$ are only  forwarded to nodes in $\Omega$.

Define a network $N$'s state to be the cartesian product of the state in all middleboxes in $N$. We say some state $s$ is reachable in $N$ if and only if there exists a schedule (given by the output of the scheduling oracle) at the end of which the network has state $s$. A subnetwork $\Omega$ of network $N$ is then closed under state if and only if there exists an equivalence relation between states in $\Omega$ and states in $N$ such that for all states reachable in $N$, the equivalent state is reachable in $\Omega$, and vice versa.

A slice is a subnetwork that is both closed under forwarding and state. Any invariant referencing only nodes and packets belonging to a slice holds in the original network if and only if it holds in the slice. Consider an invariant $I$ that is violated in some network $N$. Proving that an invariant is violated is equivalent to finding a schedule $S$ (\ie a sequence of events) which lead to the invariant being violated. Now consider $\Omega$, a slice of $N$, such that $I$ only references nodes and packets belonging to $\Omega$. Intuitively, the closure properties imply that there exists a schedule $S'$ for $\Omega$ that is equivalent to $S$. Furthermore, this equivalence implies that $S'$ also leads to $I$ being violated in $\Omega$. Finally, note that $\Omega$ is a subnetwork of $N$, and hence any schedule for $\Omega$ is also a schedule for $N$. 

\begin{figure*}[t!]
\hspace*{-0.2in}
\centering
\begin{minipage}{.33\textwidth}
    \centering
    \includegraphics[width=\textwidth]{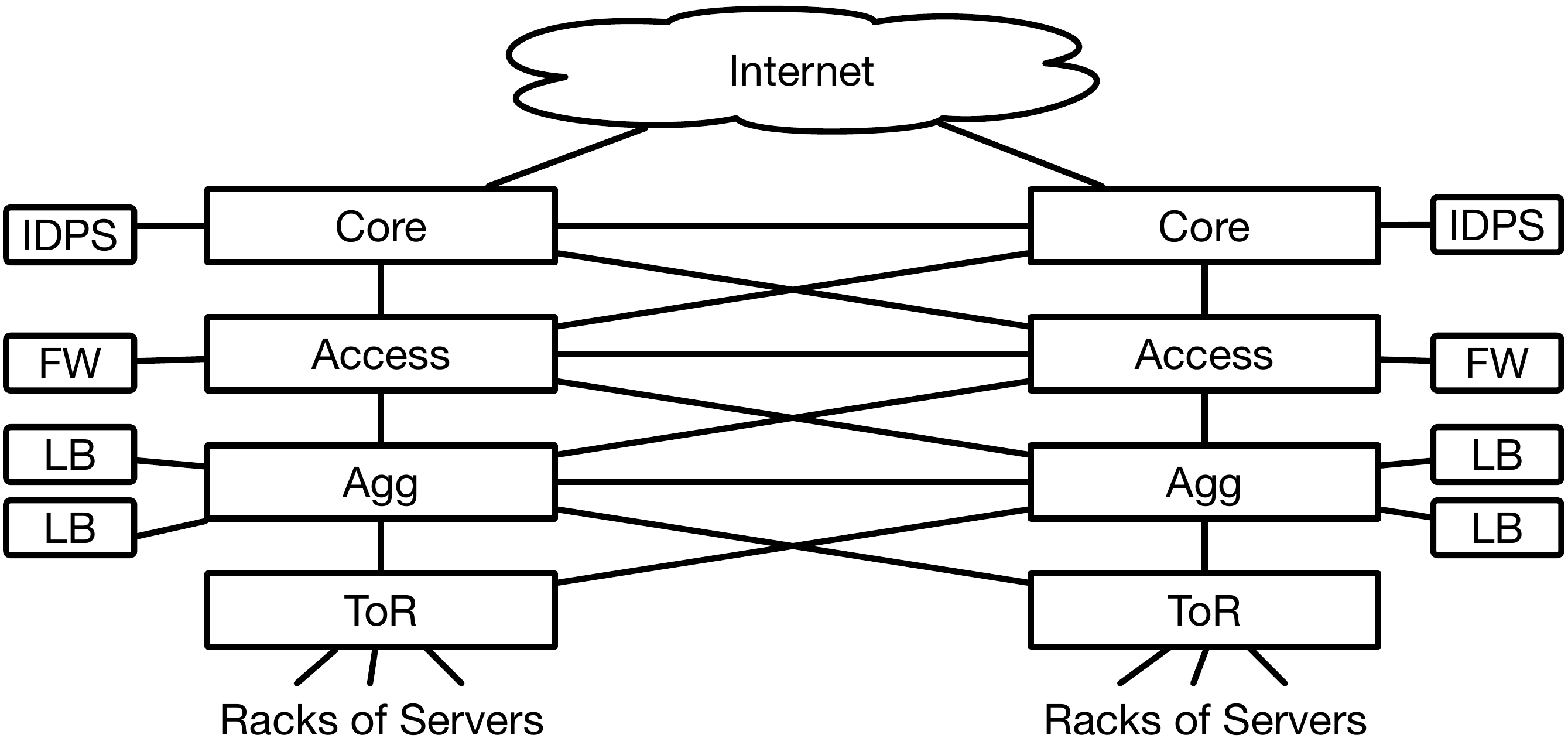}
    \caption{Topology for a datacenter network with middleboxes from~\cite{potharaju2013demystifying}. The topology contains firewalls (\textbf{FW}), load balancers (\textbf{LB}) and intrusion detection and prevention systems (\textbf{IDPS}).}
    \label{fig:real-topo}
    \vspace{-0.1in}
\end{minipage}
\hspace*{0.05in}
\begin{minipage}{.33\textwidth}
    \centering
    \includegraphics[width=\textwidth]{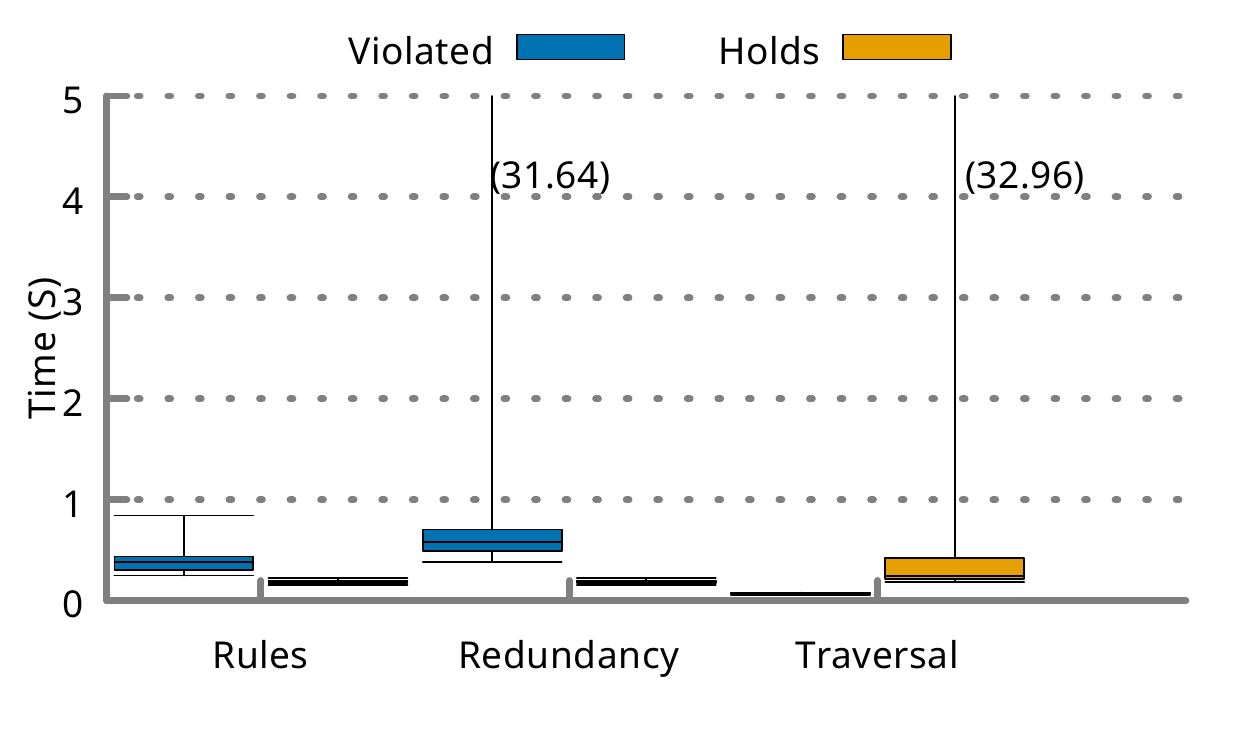}
    \caption{Time taken to verify each network invariant for scenarios in \S\ref{sec:eval:real}. We show time for checking both when invariants are violated (Violated) and verified (Holds).}
    \label{fig:imc-individual}
    \vspace{-0.1in}
\end{minipage}
\hspace*{0.05in}
\begin{minipage}{.33\textwidth}
    \includegraphics[width=\textwidth]{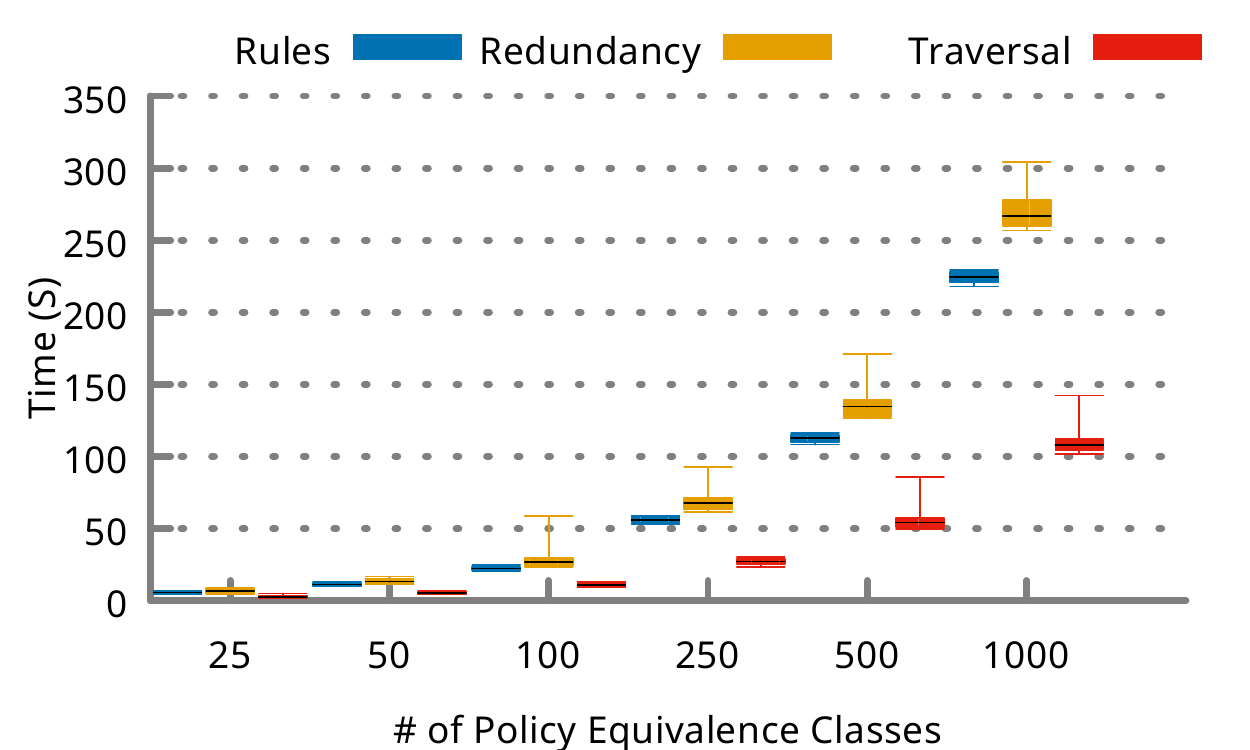}
    \caption{Time taken to verify all network invariants as a function of policy complexity for \S\ref{sec:eval:real}. The plot presents minimum, maximum, $5^{th}$, $50^{th}$ and $95{th}$ percentile time for each.}
    \label{fig:imc-all}
    \vspace{-0.1in}
\end{minipage}
\end{figure*}

\subsection{Finding Slices}
Networks with arbitrary middleboxes need not have slices smaller than the network as a whole. However, we find there is a class of networks that do have slices that do not grow with the size of the overall network.  These special networks, which we now focus on, obey the following conditions: (a) any middleboxes used in these networks be \emph{flow-parallel}, or \emph{origin-agnostic}, (b) network policies be such that we can divide hosts in the network into a set of policy equivalence classes, two hosts are in the same equivalence class if all packets sent and received by them traverse the same set of middlebox types, and are treated according to the same policy, (c) the number of policy classes be independent of the size of the network, (d) the network's forwarding graph is finite, and (e) the size of forwarding graphs is independent of network size. We define our restrictions in greater detail below.

A middlebox is \emph{flow-parallel} if middlebox state is partitioned by flows, and a flow's state is accessed only when processing that flow; \eg stateful firewalls maintain state about whether a particular flow is allowed, however this state does not affect other flows, nor is it updated by any other flow. A middlebox is \emph{origin-agnostic} if it is not flow-parallel (\ie state is shared across flows) and its behavior is agnostic to which flows (and hence hosts) instantiated the state. For example, the behavior of content-caches often does not depend on the connection that led to content being cached.

Any subnetwork that contains only \emph{flow-parallel} middleboxes and is closed under forwarding is also closed under state; the set of packets belonging to the subnetwork can be naturally mapped to a set of flows belonging to the subnetwork, and the state for those flows can only be affected by nodes in the subnetwork. Therefore, finding a slice in a network containing only flow-parallel middleboxes is equivalent to finding a subnetwork that is closed under forwarding. Therefore, verifying an invariant in a network with only \emph{flow-parallel} middleboxes only requires that we consider a subnetwork that is closed under forwarding and includes all nodes specified by the invariant. Since we assumed the size of the forwarding graph is finite and independent of network size, this means that the slices used to verify invariants in such networks are also finite.

A subnetwork that is closed under forwarding and contains only \emph{origin-agnostic} or \emph{flow-parallel} middleboxes is closed under state if and only if it includes a node from each policy class. This is because all nodes in the same policy class are equivalent for \emph{origin-agnostic} middleboxes, \ie the middlebox cannot distinguish between hosts in the same equivalence class, ensuring that the slice is closed under state. Therefore, when networks meet our requirements, all reachability invariants can be verified in network slices whose size is independent of network size.

Finally, we note that our restrictions are commonly met by deployed networks: previous studies have shown that firewalls, proxies and IDSes are the most commonly deployed middlebox types; of these firewalls are \emph{flow-parallel}, most proxies are \emph{origin-agnostic} and many IDSes are off path and do not affect reachability invariants. Furthermore, IDSes can be safely treated as \emph{origin-agnostic} middleboxes in \name, without loss in verification fidelity. For ease of management, network policy in large deployed networks is commonly expressed in terms of policy classes, and the forwarding graph is restricted for performance. Therefore, we do not believe these restrictions pose a severe challenge for \name. Finally, \name can still be used to verify moderate sized networks which violate these restrictions.
\subsection{Network Symmetry}
Slices allow us to scale verification of an individual invariant, however a single network might enforce several invariants, and the number of invariants might grow with network size. However, networks are often symmetric with respect to policy classes, \ie packets whose source and destination belong to the same policy class traverse the same sequence of middlebox types. When possible \name takes advantage of this symmetry to reduce the number of invariants to be verified. We say two invariants are symmetric when one can be transformed to another by replacing nodes with other nodes in the same policy class. If an invariant $I$ holds in a symmetric network, then so do all invariants symmetric to $I$. When networks are symmetric, \name uses this observation to divide invariants into symmetric groups and proves just one invariant in each symmetry group, allowing us to eliminate many invariant checks. 

\begin{figure*}[t!]
\hspace*{-0.2in}
\centering
\begin{minipage}{.46\textwidth}
    \centering
    \includegraphics[width=\textwidth]{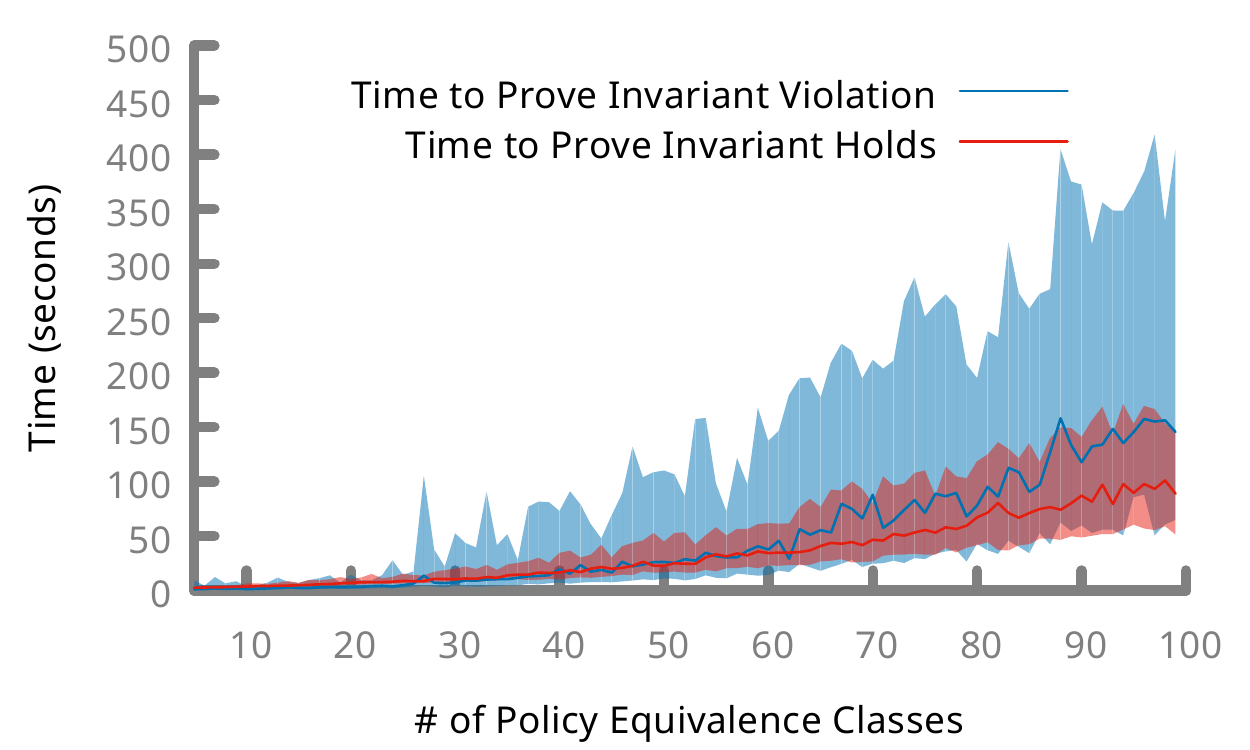}
    \caption{Time taken to verify each data isolation invariant. The shaded region represents the $5^{th}$--$95^{th}$ percentile time.}
    \label{fig:dataisolation-per-invariant}
    \vspace{-0.1in}
\end{minipage}
\hspace*{0.05in}
\begin{minipage}{.46\textwidth}
    \includegraphics[width=\textwidth]{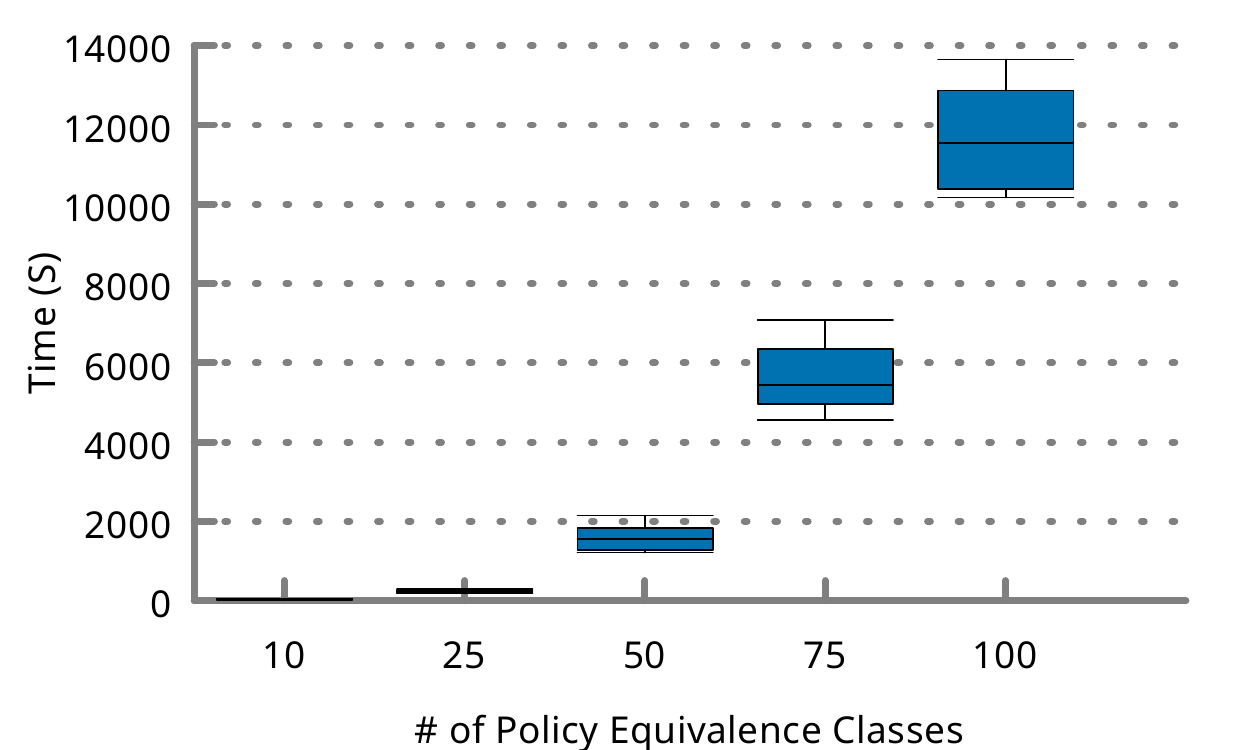}
    \caption{Time taken to verify all data isolation invariants in the network described in \S\ref{sec:eval:data}.}
    \label{fig:dataisolation-all}
\end{minipage}
\hspace*{0.05in}
\end{figure*}

\section{Evaluation}
\label{sec:eval}
\vspace{-0.09in}
To evaluate \name we first examine how it would deal with several real-world scenarios and then investigate how it scales to large networks. 
We ran our evaluation on servers running 10-core, 2.6GHz Intel Xeon processors with 256 GB of RAM. We report times taken when verification is performed using a single core. Verification can be trivially parallelized over multiple invariants. We used Z3 version 4.4.2 for our evaluation. SMT solvers rely on randomized search algorithms, and their performance can vary widely across runs. The results reported here are generated from $100$ runs of each experiment.

\subsection{Real-World Evaluation}
\label{sec:eval:real}
A previous measurement study~\cite{potharaju2013demystifying} looked at more than 10 datacenters over a 2 year period, and found that configuration bugs (in both middleboxes and networks) are a frequent cause of failure. Furthermore, the study analyzed the use of redundant middleboxes for fault tolerance, and found that redundancy failed due to misconfiguration roughly 33\% of the time. Here we show how \name can detect and prevent the three most common classes of configuration errors, including errors affecting fault tolerance. For our evaluation we use a datacenter topology (Figure~\ref{fig:real-topo}) containing $1000$ end hosts and three types of middleboxes: stateful firewalls, load balancers and intrusion detection and prevention systems (IDPSs). We use redundant instances of these middleboxes for fault tolerance. For each scenario we report time taken to verify a single invariant (Figure~\ref{fig:imc-individual}), and time taken to verify all invariants (Figure~\ref{fig:imc-all}); and show how these times grow as a function of policy complexity (as measured by the number of policy equivalence classes). Each box and whisker plot shows minimum, $5^{th}$ percentile, median, $95^{th}$ percentile and maximum time for verification.

\textbf{Incorrect Firewall Rules:} According to \cite{potharaju2013demystifying}, 70\% of all reported middlebox misconfiguration are attributed to incorrect rules installed in firewalls. To evaluate this scenario we begin by assigning each host to one of a few policy groups.\footnote{Note, policy groups are distinct from policy equivalence class; a policy group signifies how a network administrator might group hosts while configuring the network, however policy equivalence classes are assigned based on the actual network configuration.} We then add firewall rules to prevent hosts in one group from communicating with hosts in any other group. We  introduce misconfiguration by deleting a random set of these firewall rules. We use \name to identify for which hosts the desired invariant holds (\ie that hosts can only communicate with other hosts in the same group). Note that all middleboxes in this evaluation are flow-parallel, and hence the size of a slice on which invariants are verified is independent of both policy complexity and network size. In our evaluation, we found that \name correctly identified all violations, and did not report any false positives. The time to verify a single invariant is shown in Figure~\ref{fig:imc-individual} under Rules. When verifying the entire network, we only need to verify as many invariants as policy equivalence classes; hosts affected by misconfigured firewall rules fall in their own policy equivalence class, since removal of rules breaks symmetry. Figure~\ref{fig:imc-all} (Rules) shows how whole network verification time scales as a function of policy complexity.

\textbf{Misconfigured Redundant Firewalls} Redundant firewalls are often misconfigured so that they do not provide fault tolerance. To show that \name can detect such errors we took the networks used in the preceding simulations (in their properly configured state) and  introduced misconfiguration by removing rules from some of the backup firewall. In this case invariant violation would only occur when middleboxes fail. We found \name correctly identified all such violations, and we show the time taken for each invariant in Figure~\ref{fig:imc-individual} under ``Redundant'', and time taken for the whole network in Figure~\ref{fig:imc-all}. 

\textbf{Misconfigured Redundant Routing} Another way that redundancy can be rendered ineffective by misconfiguration is if routing (after failures) allows packets to bypass the middleboxes specified in the pipeline invariants. To test this we considered, for the network described above, an invariant requiring that all packet in the network traverse an IDPS before being delivered to the destination host.  We changed a randomly selected set of routing rules so that some packets would be routed around the redundant IDPS when the primary had failed. \name correctly identified all such violations, and we show times for individual and overall network verification under ``Traversal'' in Figures~\ref{fig:imc-individual} and~\ref{fig:imc-all}.

We can thus see that verification, as provided by \name, can be used to prevent many of the configuration bugs reported to affect today's production datacenters. Moreover, the verification time scales linearly with the number of policy equivalence classes (with a slope of about three invariants per second).  We now turn to more complicated invariants involving data isolation.

\subsection{Data Isolation}
\label{sec:eval:data}
Modern data centers also run storage services such as S3~\cite{awss3}, AWS Glacier~\cite{awsglacier}, and Azure Blob Store~\cite{azureblob}. These storage services must comply with legal and customer requirements~\cite{pasquier2015expressing} limiting access to this data. Operators often add caches to these services to improve performance and reduce the load on the storage servers themselves, but if these caches are misplaced or misconfigured then the access policies could be violated. \name can verify these data isolation invariants.

To evaluate this functionality, we used the topology (and correct configuration) from \S\ref{sec:eval:real} and added a few content caches by connecting them to top of rack switches. We also assume that each policy group contains separate servers with private data (only accessible within the policy group), and servers with public data (accessible by everyone). We then consider a scenario where a network administrator inserts caches to reduce load on these data servers. The content cache is configured with ACL entries\footnote{This is a common feature supported by most open source and commercial caches.} that can implement this invariant. Similar to the case above, we introduce configuration errors by deleting a random set of ACLs from the content cache and firewalls.

\begin{figure*}[t!]
\hspace*{-0.2in}
\centering
\begin{minipage}{.33\textwidth}
\hfill
    \includegraphics[width=\textwidth]{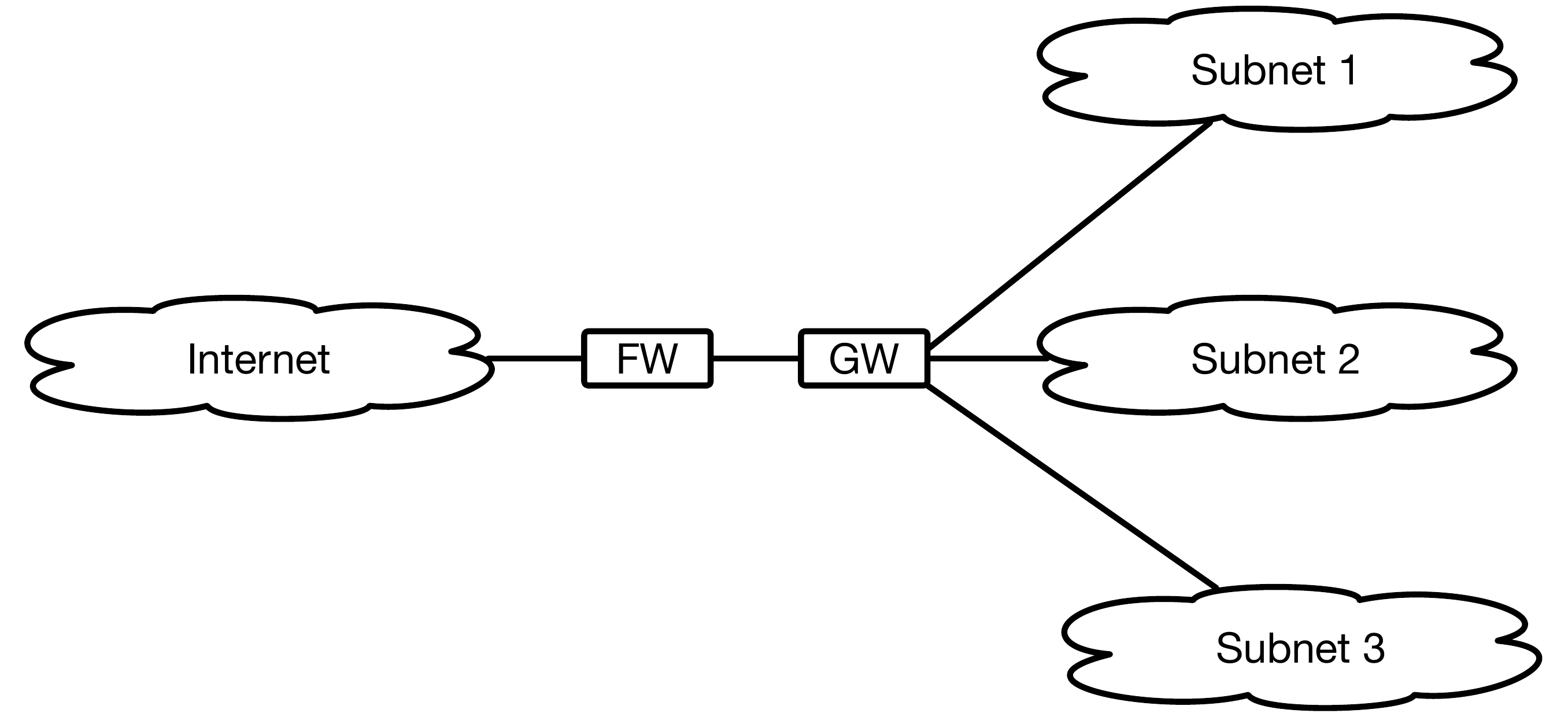}
    \caption{Topology for enterprise network used in \S\ref{sec:eval:fw}, containing a firewall (\textbf{FW}) and a gateway (\textbf{GW}).}
    \label{fig:enterprise}
\end{minipage}
\begin{minipage}{.33\textwidth}
  \centering
  \includegraphics[width=\textwidth]{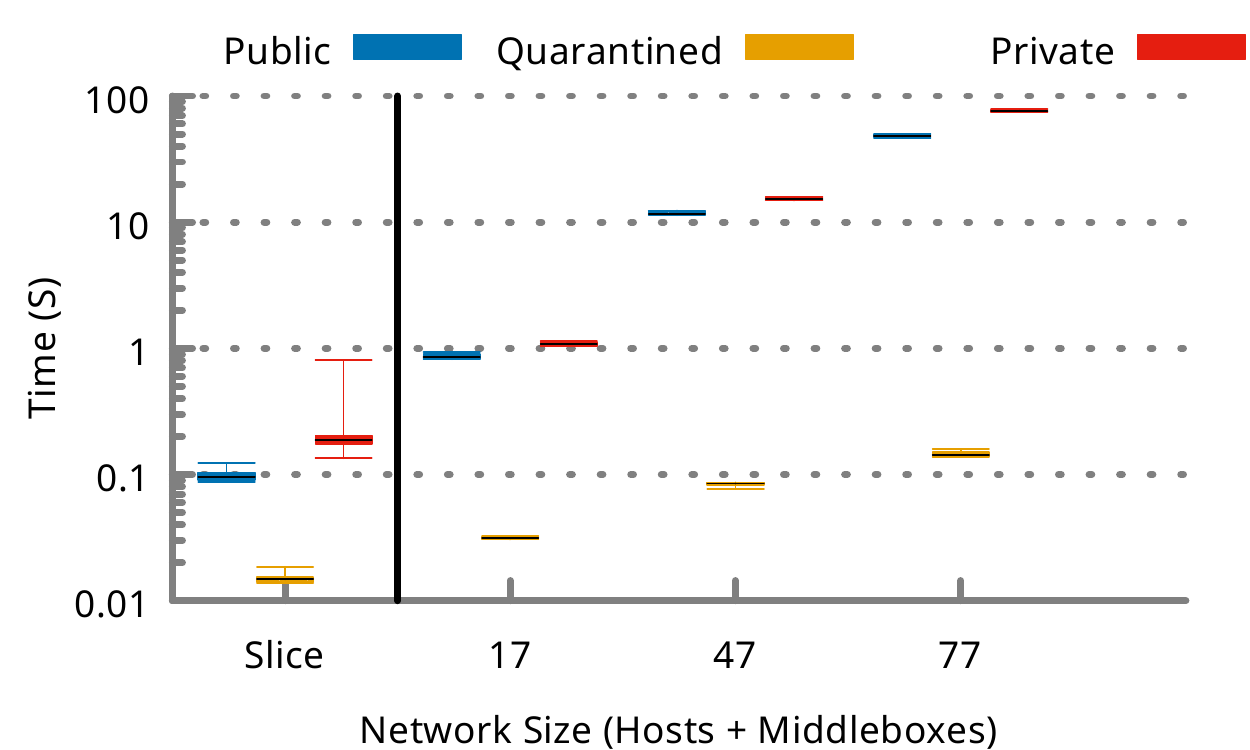}
  \caption{Distribution of verification time for each invariant in an enterprise network (\S\ref{sec:eval:fw}) with network size. The left of the vertical line shows time taken to verify a slice, which is independent of network size, the right shows time taken when slices are not used.}
  \label{fig:singlefw_pipe}
  \vspace{-0.1in}
\end{minipage}
\hspace*{0.05in}
\hspace*{0.05in}
\begin{minipage}{.33\textwidth}
    \vspace{-0.1in}
    \centering
    \includegraphics[width=\textwidth]{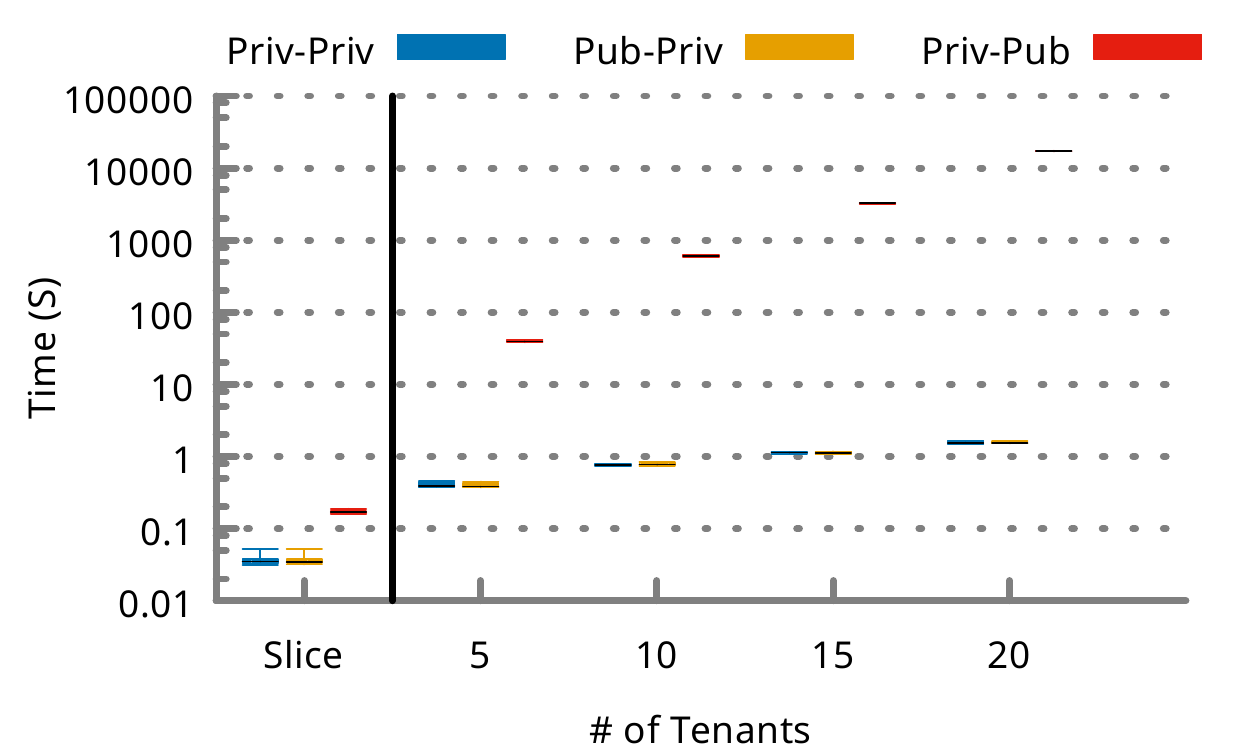}
    \vspace{-0.1in}
    \caption{Average verification time for each invariant in a multi-tenant datacenter (\S\ref{sec:eval:dc}) as a function of number of tenants. Each tenant has 10 hosts. The left of the vertical line shows time taken to verify a slice, which is independent of the number of tenants.}
    \label{fig:mtdc}
\end{minipage}
\hspace*{0.05in}
\end{figure*}
We use \name to verify data isolation invariants in this network (\ie ensure that private data is only accessible from within the same policy group, and public data is accessible from everywhere). \name correctly detects invariant violations, and does not report any false positives. Content caches are origin agnostic, and hence the size of a slice used to verify these invariants depends on policy complexity. Figure~\ref{fig:dataisolation-per-invariant} shows how time taken for verifying each invariant varies with the number of policy equivalence classes. In a network with $100$ different policy equivalence classes, verification takes less than $4$ minutes on average. Also observe that the variance for verifying a single invariant grows with the size of slices used. This shows one of the reasons why the ability to use slices and minimize the size of the network on which an invariant is verified is important. Figure~\ref{fig:dataisolation-all} shows time taken to verify the entire network as we increase the number of policy equivalence classes.

\subsection{Other Network Scenarios}
We next apply \name to several other scenarios that illustrate the value of slicing (and symmetry) in reducing verification time.

\subsubsection{Enterprise Network with Firewall}
\label{sec:eval:fw}

First, we consider a typical enterprise or university network protected by a stateful firewall, shown in Figure~\ref{fig:enterprise}. The network interconnects three types of hosts:
\begin{asparaenum}
\item Hosts in \emph{public} subnets should be allowed to both initiate and accept connections with the outside world. 
\item Hosts in \emph{private} subnets should be flow-isolated (\ie allowed to  initiate connections to the outside world, but never accept incoming connections).
\item Hosts in \emph{quarantined} subnets should be node-isolated (\ie not allowed to communicate with the outside world).
\end{asparaenum}
We vary the number of subnets keeping the proportion of subnet types fixed; a third of the subnets are public, a third are private and a third are quarantined.

We configure the firewall so as to enforce the target invariants correctly: with two rules denying access (in either direction) for each quarantined subnet, plus one rule denying inbound connections for each private subnet. The results we present below are for the case where all the target invariants hold. Since this network only contains a firewall, using slices we can verify invariants on a slice whose size is independent of network size and policy complexity. We can also leverage the symmetry in both network and policy to reduce the number of invariants that need to be verified for the network. In contrast, when slices and symmetry are not used, the model for verifying each invariant grows as the size of the network, and we have to verify many more invariants. In Figure~\ref{fig:singlefw_pipe} we show time taken to verify the invariant using slices (Slice) and how verification time varies with network size when slices are not used. 

\begin{figure*}[tb]
\subfigure[]{
    \centering
        \includegraphics[width=0.31\textwidth]{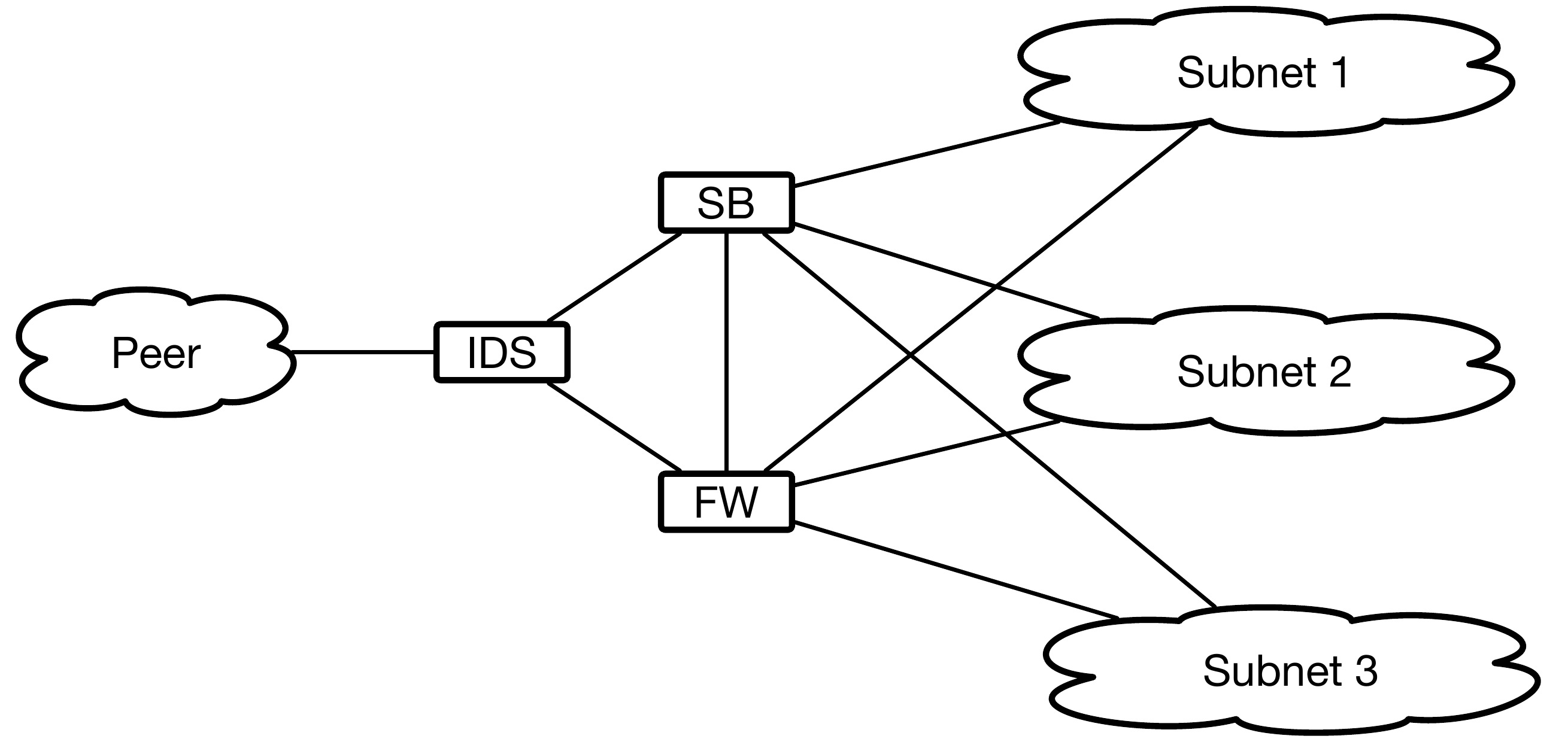}
        \label{fig:wan-ids-scrub}
}
\hfill
\subfigure[]{
    \centering
    \includegraphics[width=0.31\textwidth]{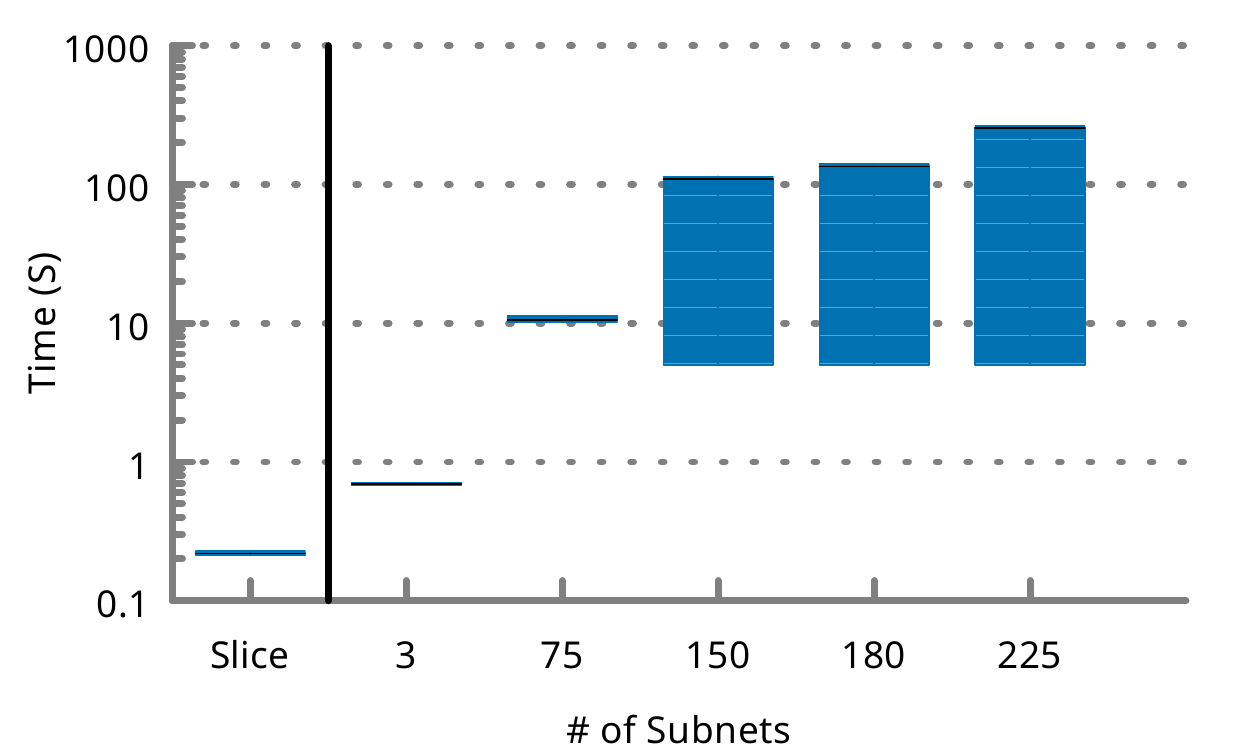}
    \label{fig:ids-vary-int-1s}
}
\hfill
\subfigure[]{
    \centering
    \includegraphics[width=0.31\textwidth]{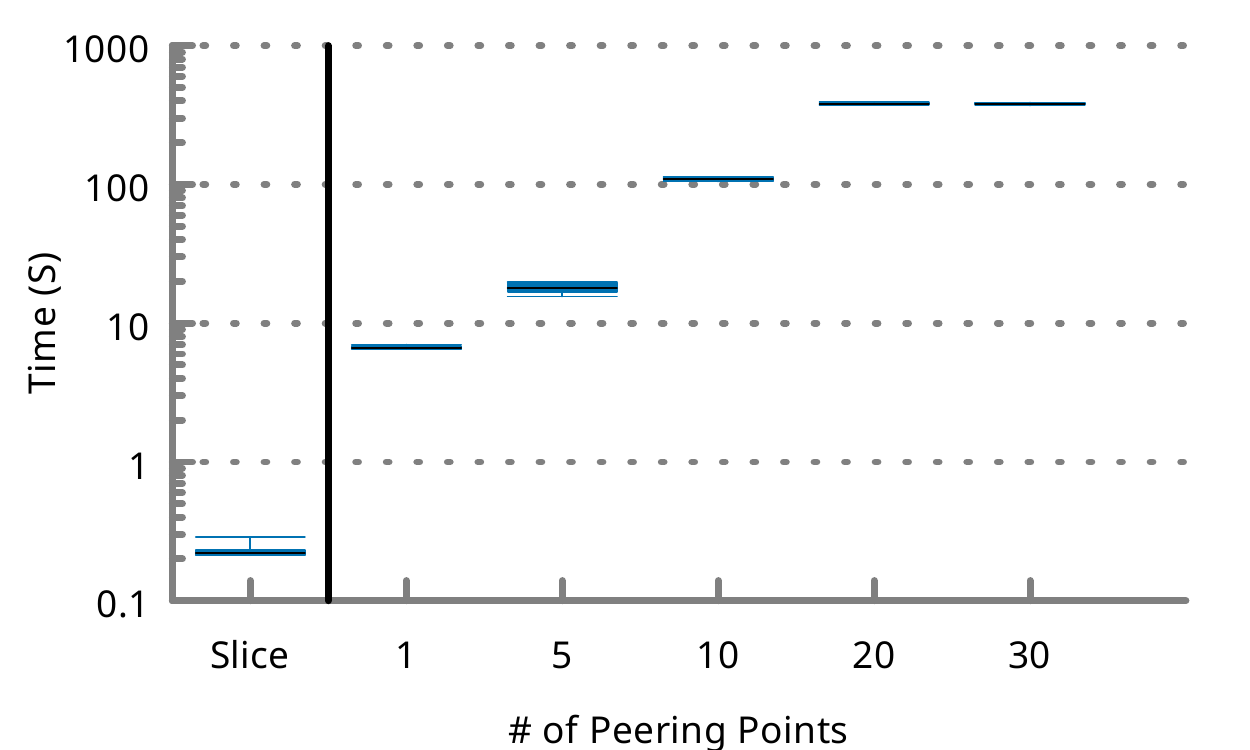}
    \label{fig:ids-vary-ext-1s}
}
\caption{\subref{fig:wan-ids-scrub} shows the pipeline at each peering point for an ISP; \subref{fig:ids-vary-int-1s} distribution of time to verify each invariant given this pipeline when the ISP peers with other networks at $5$ locations; \subref{fig:ids-vary-ext-1s} average time to verify each invariant when the ISP has $75$ subnets. In both cases, to the left of the black line we show time to verify on a slice (which is independent of network size) and vary sizes to the right.}
    \vspace{-0.2in}
\end{figure*}

\subsubsection{Multi-Tenant Datacenter}
\label{sec:eval:dc}
Next, we consider how \name can be used by a cloud provider (\eg Amazon) to verify isolation in a multi-tenant datacenter. We assume that the datacenter implements the Amazon EC2 Security Groups model~\cite{awssg}. For our test we considered a datacenter with $600$ physical servers (which each run a virtual switch) and $210$ physical switches (which implement equal cost multi-path routing). Tenants launch virtual machines (VMs), which are run on physical servers and connect to the network through virtual switches. Each virtual switch acts as a stateful firewall, and defaults to denying all traffic (\ie packets not specifically allowed by an ACL are dropped). To scale policy enforcement, VMs are organized in security groups with associated accept/deny rules. For our evaluation, we considered a case where each tenant organizes their VMs into two security groups:
\begin{asparaenum}
\item VMs that belong to the \emph{public} security group are allowed to accept connections from any VMs. 
\item VMs that belong to the \emph{private} security group are flow-isolated (\ie they can initiate connections to other tenants' VMs,  but can only accept connections from this tenant's public and private VMs). 
\end{asparaenum}

We also assume that firewall configuration is specified in terms of security groups (\ie on receiving a packet the firewall computes the security group to which the sender and receiver belong and applies ACLs appropriately). For this evaluation, we configured the network to correctly enforce tenant policies. We added two ACL rules for each tenant's public security group allowing incoming and outgoing packets to anyone, while we added three rules for private security groups; two allowing incoming and outgoing traffic from the tenant's VM, and one allowing outgoing traffic to other tenants. For our evaluation we consider a case where each tenant has $10$ VMs, $5$ public and $5$ private, which are spread across physical servers. These rules result in flow-parallel middleboxes, so we can  use fixed size slices to verify each invariant. The number of invariants that need to be verified grow as a function of the number of tenants. In Figure~\ref{fig:mtdc} we show time taken to verify one instance of the invariant when slices are used (Slice) and how verification time varies with network size when slices are not used. The invariants checked are: (a)  private hosts in one group cannot reach private hosts in another group (Priv-Priv), (b) public hosts in one group cannot reach private hosts in another group (Priv-Pub), and (c) private hosts in one group \emph{can} reach public hosts in another.

\subsubsection{ISP with Intrusion Detection}
\label{sec:eval:ids}
Finally, we consider an Internet Service Provider (ISP) that implements an intrusion detection system (IDS). We model our network on the SWITCHlan backbone~\cite{switchback}, and assume that there is an IDS box and a stateful firewall at each peering point  (Figure~\ref{fig:wan-ids-scrub}). The ISP contains public, private and quarantined subnets (with policies as defined in \secref{sec:eval:fw}) and the stateful firewalls enforce the corresponding invariants. Additionally, each IDS performs lightweight monitoring (\eg based on packet or byte counters) and checks whether a particular destination prefix (\eg a customer of the ISP) might be under attack; if so, all traffic to this prefix is rerouted to a scrubbing box that performs more heavyweight analysis, discards any part of the traffic that it identifies as ``attack traffic,'' and forwards the rest to the intended destination. This combination of multiple lightweight IDS boxes and one (or a few) centralized scrubbing boxes is standard practice in ISPs that offer attack protection to their customers.\footnote{This setup is preferred to installing a separate scrubbing box at each peering point because of the high cost of these boxes, which can amount to several million dollars for a warranteed period of $3$ years.} 

To enforce the target invariants (for public, private, and quarantined subnets) correctly, all inbound traffic must go through at least one stateful firewall. We consider a misconfiguration where traffic rerouted by a given IDS box to the scrubbing box bypasses all stateful firewalls. As a result, any part of this rerouted traffic that is \emph{not} discarded by the scrubbing box can reach private or quarantined subnets, violating the (simple or flow-) isolation of the corresponding hosts.

When verifying invariants in a slice we again take advantage of the fact that firewalls and IDSes are flow-parallel.\footnote{While IDSes in general might not be flow-parallel, the specific IDS used here is flow-parallel with respect to a slice.} For each subnet, we can verify invariants in a slice containing a peering point, a host from the subnet, the appropriate firewall, IDS and a scrubber. Furthermore, since all subnets belong to one of three policy equivalence classes, and the network is symmetric, we only need run verification on three slices. 

We begin by evaluating a case where the ISP, similar to the SWITCHlan backbone has $5$ peering points with other networks. We measure verification time as we vary the number of subnets (Figure~\ref{fig:ids-vary-int-1s}), and report time taken, on average, to verify each invariant. When slices are used, the median time for verifying an invariant is $0.21$ seconds, by contrast when verification is performed on the entire network, a network with $250$ subnets takes approximately $6$ minutes to verify. Furthermore, when verifying all invariants, only $3$ slices need to be verified when we account for symmetry, otherwise the number of invariants verified grows with network size.

In Figure~\ref{fig:ids-vary-ext-1s} we hold the number of subnets constant (at $75$) and show verification time as we vary the number of peering points. In this case the complexity of verifying the entire network grows more quickly (because the IDS model is more complex leading to a larger increase in problem size). In this case, verifying correctness for a network with $50$ peering points, when verification is performed on the whole entire network, takes approximately $10$ minutes. Hence, being able to verify slices and use symmetry is crucial when verifying such networks.

\section{Related Work}
\label{sec:related}
Next, we discuss related work in network verification and formal methods. 

\paragraphb{Testing Networks with Middleboxes}
The work most closely related to us is Buzz~\cite{fayaz2014buzz}, which uses symbolic execution to generate sequences of packets that can be used to test whether a network enforces an invariant. Testing, as provided by Buzz, is complimentary to verification. Our verification process does not require sending traffic through the network, and hence provides a non-disruptive mechanism for ensuring that changes to a network (\ie changing middlebox or routing configuration, adding new types of middleboxes, etc.) do not result in invariant violation. Verification is also useful when initially designing a network, since designs can be evaluated to ensure they uphold desirable invariants. However, as we have noted in \S\ref{sec:system:limitations}, our verification results hold if and only if middlebox implementations are correct, \ie packets are correctly classified, etc. Combining a verification tool like \name with a testing tool such as Buzz allows us to circumvent this problem, when possible (\ie when the network is not heavily utilized, or when adding new types of middleboxes), Buzz can be used to test if invariants hold. This is similar to the complimentary use of verification and testing in software development today. More specifically, it is almost identical to the relationship between ATPG (testing) and HSA (verification).

Beyond the difference of purpose, there are some other crucial difference between Buzz and \name: (a) Buzz's middlebox models are context specific, and must be specialized for each network, while \name's models are more general and designed to be reused across networks; (b) Buzz's testing does not consider the effect of middlebox failure, while our approach can be used to verify that invariants hold despite network failures; and (c) scaling to large networks with slicing is one of our major contributions. Buzz can scale to networks with 100s of nodes before running into scaling limits, while in many cases we can scale to arbitrary sized networks. We believe slicing can also be used by Buzz to improve scaling.

SymNet~\cite{stoenescu2013symnet} has also suggested the need to extend these mechanisms to handle mutable datapath elements. SymNet uses
symbolic execution to check reachability properties, however their technique is only applicable when state is not shared across a flow and was only applied to cases where the middlebox can punch holes, but do no more. They can therefore only deal with a few kinds of middleboxes, and it is unclear if their technique can be extended to scalably handle other middlebox types.

\paragraphb{Verifying Forwarding Rules}
Recent efforts in network verification~\cite{mai2011debugging,nsdi:CaniniVPKR12,kazemian2012header,khurshid2012veriflow,Verificare,FMACAD:SNM13,anderson2014netkat,foster2015coalgebraic}
have focused on verifying the network dataplane by analyzing forwarding tables. Some of these tools including HSA~\cite{kazemian2013real},
Libra~\cite{zeng2014libra} and VeriFlow~\cite{khurshid2012veriflow} have also developed algorithms to perform near real-time verification of simple 
properties such as loop-freedom and the absence of blackholes. Recent work~\cite{symmetryandsurgery} has also shown how techniques similar to slicing can be used to scale these techniques. Our approach to slicing generalizes this work by accounting for state. While these techniques are well suited for checking networks with static data planes they are insufficient for dynamic datapaths. 

\paragraphb{Verifying Network Updates}
Another line of network verification research has focused on verification during configuration updates. This line of work can
be used to verify the consistency of routing tables generated by SDN controllers~\cite{KattaRW13,VanbeverRBFR13}.
Recent efforts~\cite{tr:MSRSegguru} have generalized these mechanisms and can be used to determine what parts of the
configuration are affected by an update, and verify invariants on this subset of the configuration. This line of work
has been restricted to analyzing policy updates performed by the control plane and does not address dynamic data plane elements
where state updates are more frequent.

\paragraphb{Verifying Network Applications}
Other work has looked at verifying the correctness of control and data plane applications. NICE~\cite{nsdi:CaniniVPKR12} proposed using static analysis to verify the
correctness of controller programs. Later extensions including~\cite{conext:uzniarPCVK12} have looked at improving the accuracy of NICE using
concolic testing~\cite{sen2006cute} at the cost of completeness. More recently, Vericon~\cite{ball2014vericon} has looked at sound verification of a restricted class of controllers.

Recent work~\cite{dobrescu2014software} has also looked at using symbolic execution to prove properties for programmable datapaths (middleboxes). This work in particular looked at verifying
bounded execution, crash freedom and that certain packets are filtered for stateless or simple stateful middleboxes written as pipelines and meeting certain criterion. The verification
technique does not scale to middleboxes like content caches which maintain arbitrary state.

\paragraphb{Finite State Model Checking}
Finite state model checking has been applied to check the correctness of many hardware and software based systems~\cite{book:CGD01}. Here the behavior
of a system is specified as a transition relation between finite state and a checker can verify that all reachable states from a starting configuration
are safe (\ie do not cause any invariant violation). Tools such as NICE~\cite{nsdi:CaniniVPKR12}, HSA~\cite{kazemian2012header} and others~\cite{CAV:SosnovichGN13}
rely on this technique. However these techniques scale exponentially with the number of states and for even moderately large problems one must choose between being
able to verify in reasonable time and completeness. Our use of SMT solvers allows us to reason about potentially infinite state and 
our use of simple logic allows verification to terminate in a decidable manner for practical networks.

\paragraphb{Language Abstractions}
Several recent works in software-defined networking~\cite{IEEECOMM:FosterGRSFKMRRSWH13,sigcomm:VoellmyWYFH13,PLDI:GuhaRF13, FlowLog, NLOG} have
proposed the use of verification friendly languages for controllers. One could similarly extend this concept to provide a verification friendly
data plane language however our approach is orthogonal to such a development: we aim at proving network wide properties rather than properties for individual middleboxes.

\vspace{-0.1in}
\section{Conclusion}
\label{sec:conclusion}

We started this work aiming to extend the benefits of verification to networks with middleboxes. In building \name, we had three significant
realizations: first, we should separate the classification and forwarding behavior of middleboxes, and only abstractly model classification. Second,
scaling verification requires us to take advantage of how state is partitioned by middleboxes, and of the symmetry in the network topology and policies. Lastly,
verifying invariants in the presence of failures is essential to making middlebox verification useful in existing networks. The combination of the first two steps
enables verification of reachability invariants on extremely large networks; while the last allows us to verify most previously reported configuration bugs. 
 Since our results depend on network symmetry they do not entirely apply to random networks, and scalability for general networks remains an important open problem. However, we note that in other domains where verification has been successfully employed, scaling has been achieved by taking advantage of the problem structure, and we believe our work on slices is analogous to this work.





\balance
\bibliographystyle{acm}
\bibliography{bibs}
\end{document}